\documentclass[aps,pra,twocolumn,showpacs,superscriptaddress,groupedaddress]{revtex4}
\usepackage{graphicx} 
\usepackage{amsmath}

\begin{document}

\title{Temperature Measurement via Time Crystal Frequencies in One-Dimensional Quantum Droplets}
\author{Saurab Das}
\affiliation{Indian Institute of Information Technology Vadodara, Gujarat, India 382 028}
\author{Jagnyaseni Jogania}
\author{Jayanta Bera}
\affiliation{C. V. Raman Global University, Bhubaneswar, Odisha 752 054, India}
\author{Ajay Nath}
\affiliation{Indian Institute of Information Technology Vadodara, Gujarat, India 382 028}

\begin{abstract}
We propose a novel approach for temperature measurement in ultracold quantum systems by analyzing the frequency of emergent time crystals in one-dimensional (1D) quantum droplets. The system under investigation comprises a binary Bose-Einstein condensate confined within a driven quasi-periodic optical lattice (QOL), incorporating repulsive mean-field and attractive beyond-mean-field interactions. Employing the extended 1D Gross–Pitaevskii equation, we derive an analytical form of the wavefunction and systematically explore droplet dynamics under three distinct driving protocols: (i) increasing QOL depth at fixed frequency, (ii) varying driving frequency at constant depth, and (iii) sinusoidal modulation of the lattice depth. Fourier analysis of the density oscillations reveals the formation of time crystalline states with harmonic frequency components. Crucially, we establish a correlation between the time crystal frequency and the system’s temperature, showing that variations in driving frequency induce oscillatory behavior in the droplet’s effective negative temperature. A comprehensive numerical stability analysis confirms the robustness of the time crystalline states, highlighting their potential observability in experiments. These findings open a new pathway for probing thermodynamic properties in quantum many-body systems through dynamical time crystal signatures.
\end{abstract}

\maketitle

\section{Introduction} Quantum droplets (QDs) represent an ultradilute quantum liquid state that emerges in ultracold atomic systems, exhibiting characteristics distinct from classical liquids \cite{Luo}. These droplets are typically realized in binary Bose–Einstein condensate (BEC) mixtures, where mean-field atom–atom interactions are counterbalanced by beyond-mean-field (BMF) effects, primarily due to the Lee–Huang–Yang (LHY) quantum fluctuation correction \cite{Petrov}. Initially predicted in three-dimensional Bose gas mixtures \cite{Petrov1}, QDs have subsequently been realized in one-dimensional (1D) binary BECs, where repulsive two-body interactions are stabilized by attractive LHY-induced fluctuations \cite{Hertkorn}. The confinement of these systems in reduced dimensions enhances interaction effects, making ultracold atomic gases a versatile platform for studying strongly correlated quantum liquids \cite{Giamarchi}. Earlier, Bulgac proposed the existence of “boselets,” droplet-like states stabilized by the interplay of two- and three-body interactions—alongside theoretical analogues for fermionic systems, such as “fermilets” and “ferbolets” \cite{Bulgac, Bedaque}. Recent experimental progress in the preparation, control, and detection of ultracold atomic droplets \cite{Barbut, Schmitt, Cabrera, Wenzel, Semeghini, Ferioli} has significantly advanced the field, sparking renewed interest in the underlying many-body physics \cite{Parisi, Astrakharchik, Hu, Rosi, Zin, Bhatia, Ivan, Maitri1, Nie, Cui, Sekino, Das}. The study of QDs now encompasses a broad spectrum of phenomena, including supersolid phases \cite{Parit, Mukherjee}, beyond-mean-field dimensional crossovers \cite{Lavoine, Zin1}, vortex formation \cite{Zhang}, droplet-to-soliton transitions at negative temperatures \cite{Maitri2}, dark quantum droplets \cite{Edmonds}, and enhanced mobility under lattice modulation \cite{Kartashov}. QDs have been extensively studied across diverse physical settings, unveiling phenomena such as pattern formation,  nonlinear management effects, stable rotating states, multipole and vortex configurations, and interactions within confined potentials \cite{Dong1,Chen1,Liu1,Liu2,Liu3,Dong2,Dong3,Dong4,MR1}. These investigations have significantly advanced our understanding of QD formation, stability, and dynamical evolution. Despite these advances, the generation of time crystals in quantum droplets and their potential correlation with system temperature remains largely unexplored. This unexplored intersection presents an exciting avenue for both theoretical modeling and experimental realization.

In this work, we investigate the emergence of time crystals within QDs and establish a novel link between their dynamical behavior and the negative temperature regime of the system. Both time crystals \cite{Sacha,Zhang1} and negative temperature states \cite{Braun,Carr} have garnered significant attention in the realm of ultracold atomic physics due to their fundamental implications for nonequilibrium thermodynamics and quantum many-body dynamics. Time crystals, originally conceptualized by Wilczek, arise through the spontaneous breaking of time-translation symmetry in the ground state of a quantum many-body system, mirroring the formation of spatial crystals through the breaking of spatial translational symmetry \cite{Wilczek}. Subsequent theoretical and experimental advancements revealed that such symmetry breaking can also manifest in periodically driven systems, giving rise to discrete time crystals whose oscillation period is an integer multiple of the external driving period \cite{Sacha1}. These discrete time crystals are characterized by their robustness to perturbations and long-lived coherent oscillations that persist in the thermodynamic limit \cite{Giergiel,Golletz}. Simultaneously, negative temperature states—where entropy decreases with increasing energy—have emerged as a cornerstone in the understanding of inverted population systems, provided the energy spectrum is bounded from above \cite{Ramsey}. Although initially controversial due to competing thermodynamic interpretations \cite{Dunkel,Calabrese}, negative temperatures have now been successfully realized in ultracold atomic gases and validated across a broad range of physical platforms \cite{Braun,Baldovin,Abraham}. These states have been investigated in the context of quantum heat engines \cite{Wang}, early-universe thermodynamics \cite{Vieira}, refrigeration of quantum bits \cite{Damas}, and optical lattice systems, including bi-periodic configurations (BOL) \cite{Kundu} and quantum batteries \cite{Struchtrup}. In particular, recent findings have shown that QDs can exist in negative temperature regimes, exhibiting unique nonlinear structures such as droplet-to-soliton crossovers when confined in BOLs and expulsive-BOL traps \cite{Maitri2}. Despite these advancements, the intricate interplay between time crystal formation and the thermodynamic characteristics of negative temperature QDs has remained largely unexplored. This study addresses this gap by demonstrating that the frequency of time crystalline oscillations in QDs can serve as a sensitive probe of system temperature, particularly in regimes exhibiting negative absolute temperatures. Our findings thus open a new avenue for experimentally accessible temperature diagnostics in strongly correlated quantum fluids and contribute to a deeper understanding of time-dependent many-body quantum phenomena.

Here, we present an analytical framework for the generation of time crystals and the exploration of QD dynamics by obtaining exact solutions to the one-dimensional (1D) extended Gross–Pitaevskii equation (eGPE) under a driven quasi-periodic optical lattice (QOL) at negative temperatures. Our approach captures the interplay between repulsive cubic effective mean-field (EMF) interactions and attractive quadratic beyond-mean-field (BMF) interactions arising from the Lee-Huang-Yang correction, providing a physically realizable platform for probing QD dynamics in regimes where negative temperature states can be experimentally accessed \cite{Kundu}. The analytical wavefunction is derived by solving the 1D eGPE for a binary Bose-Einstein condensate (BEC) confined within a modulated QOL. We analyze droplet dynamics under three distinct driving conditions: (i) linearly increasing QOL depth at fixed driving frequency, (ii) linearly varying driving frequency at fixed QOL depth, and (iii) sinusoidally modulated QOL depth at constant frequency. Fast Fourier Transform (FFT) analysis of the resulting condensate density oscillations reveals the presence of harmonic modes, confirming the emergence of time-crystalline behavior in the system. Notably, we establish a non-trivial correlation between the time crystal frequency and the system’s thermodynamic state: increasing frequency induces oscillatory variations in the droplet’s negative temperature. This result offers a novel perspective on temperature diagnostics in nonequilibrium quantum systems. Furthermore, to validate the robustness of our solutions, we conduct numerical simulations using the split-step Fourier transform (SSFT) method. The time evolution of the condensate density is examined under the influence of both the modulated QOL and externally introduced white noise perturbations. Our stability analysis demonstrates that the standard deviation in the condensate density remains below $2\%$ of its maximum value, indicating strong agreement between analytical and numerical results and confirming the feasibility of experimental implementation.

This study begins by presenting the theoretical foundation for modeling QDs, alongside the development of an analytical framework in Section II. In Section III, this model is employed to explore the modifications in QD states under driven quasi-periodic optical lattice (QOL) trapping potentials, in contrast to free-space conditions, with particular emphasis on the emergence of time crystals. Section IV focuses on analyzing the system’s temperature across different driving scenarios, establishing a correlation between the frequency of the generated time crystals and the negative temperature regime of the system. Section V presents numerical simulations to assess the stability of the QD solutions under time-dependent QOL potential depths. Finally, Section VI summarizes the main findings and outlines potential avenues for future research.

\section{Methods}
We model our system by considering a one-dimensional homonuclear binary BEC mixture consisting of two distinct hyperfine states of $^{39}$K \cite{Semeghini}. The binary BEC states are assumed to be symmetric, represented as $\psi_{1}=\psi_{2}=c_{0} \psi$, with equal atom numbers $N_{1}$=$N_{2}$=$N$, and identical masses. The intra-atomic coupling constants are taken to be equal, $g_{\uparrow \uparrow} = g_{\downarrow \downarrow}\equiv g (>0)$, defined as $2 \hbar^{2} a_{s}(x, t)/(m a^{2}_{\perp})$, and a negative inter-atomic coupling constant $ g_{\uparrow \downarrow} (<0)$, leading to a droplet regime characterized by $ \delta g = g_{12} + g > 0$  \cite{Astrakharchik}. Under these assumptions, the binary BEC mixture can be effectively modeled as a simplified single-component, dimensionless 1D eGPE that includes first-order LHY quantum corrections \cite{Petrov, Astrakharchik}:
\begin{eqnarray}\label{eq:QD1}
i\frac{ \partial \psi}{\partial t} =- \frac{1}{2} \frac{ \partial^2 \psi}{\partial x^2} -  \gamma_1(x,t) |\psi|\psi + \gamma_2(x,t) |\psi|^2\psi + V(x,t) \psi.  
\end{eqnarray}
In this framework, the wavefunction, length, and time are expressed in units of $ (2 \sqrt{g(t)})^{3/2}/(\pi \xi(2|\delta g(t)|)^{3/4}$, $\xi$, and $\hbar^{2}/m \xi^{2}$  respectively with $\xi=\pi \hbar^{2} \sqrt{|g(t)|}/(m g(t) \sqrt{2}g(t))$ is the healing length of the system \cite{Astrakharchik}. Here, the functions $\gamma_{1}(x,t) = = (\sqrt{2m}/ \pi \hbar) g(t)^{3/2}$ $= (\sqrt{2m}/\pi \hbar) [\sqrt(g_{11}(x,t)g_{22}(x,t))]^{3/2}$ and $\gamma_{2}(x,t) = [g_{12}(t)+\sqrt(g_{11}(x, t)g_{22}(x, t))] $ with $g_{ii} = 4 \pi a_{ii}/m_{i}$ and $g_{12} = 2 \pi a_{12}/m_{r} , m_{r}=m_{1}m_{2}/(m_{1}+m_{2})$, are non-zero, represent the space- and time-dependent coupling strengths of the BMF and EMF interactions in the binary BEC mixture, respectively. The quadratic nonlinearity in the equation accounts for the attractive nature of the 1D LHY correction, while the cubic term represents the conventional mean-field repulsion, both of which are essential for realizing the droplet state.

We consider the form of external confinement to achieve the negative temperature regime as \cite{Braun,Kundu}:
\begin{eqnarray}\label{eq:QD3}
\hspace{-0.4 cm} V(x,t) \simeq  V_{1}(t) \cos[2 k (x-v t)] 
+ \left[V_{2}(t) + \mu \beta(t) \right] \cos[ k (x-v t)],
\end{eqnarray}
with $V_{1}(t)=-\frac{\beta^{2}(t) k^{2}}{16}, \;\;V_{2}(t)=\frac{\beta(t) k^{2}}{4},$ represents the potential depths of the QOL trap typically expressed in terms of recoil energy: $E_{R}=\frac{2\pi^{2}\hbar^{2}}{m \lambda^{2}}$ \cite{Windpassinger}. Experimentally, these QOL potential depths can be adjusted by varying the laser wavelength ($\lambda$) and the mass of the BEC atoms. In this, the frequencies of the two laser beams are commensurate, with $k = 2 \pi a_{\perp}/ \lambda$ representing the scaled lattice wave vector with $a_{\perp}=\sqrt{\hbar/m \omega_{\perp}}$.  Here, $\hbar$, $\omega_{\perp}$, and $m$ represents the reduced Planck's constant, transverse oscillator frequency, and mass of binary BEC atoms, respectively and $v>0$ is a positive constant. The chosen trap in equation (\ref{eq:QD3}) is approximated form of multi-color optical lattice $V(x,t)=V_{1}(t)\cos[2 k (x-v t)] + V_{2}(t) \cos[ k (x-v t)] + \frac{1}{2} \mu \times exp[2 \beta(t) cos[ k (x-v t)]],$ with $-1<\beta(t) <1$. It is important to note that as $ \beta(t) \rightarrow 0 $, the term $ \cos[k(x - v t)] $ becomes dominant over $ \cos[2k(x - v t)] $, effectively reducing the potential to a monotonic, single-frequency periodic lattice. Consequently, the fundamental spatial driving frequencies associated with the chosen QOL trap are $ f = kv/2\pi $ for the $ \cos[k(x - v t)] $ component and $ 2f $ for the $ \cos[2k(x - v t)] $ component.

For solving equation (\ref{eq:QD1}), our aim is to connect it with a solvable ordinary differential equation of form:
\begin{equation}\label{eq:QD4}
-U_{X X}-G_1\mid U(X)\mid U +G_2\mid U(X)\mid^{2} U=\mu F. 
\end{equation}
by choosing the following ansatz solution: 
\begin{equation}\label{eq:QD5}
\psi(x,t)= A(x,t)U[X(x,t)]e^{i\phi(x,t)},
\end{equation}
with following constraints on the the forms of the amplitude [$A(x,t)$], phase [$\phi(x,t)$], and traveling coordinate [X(x,t)]:
\begin{figure*}[t]
\centering
\includegraphics[width=\linewidth]{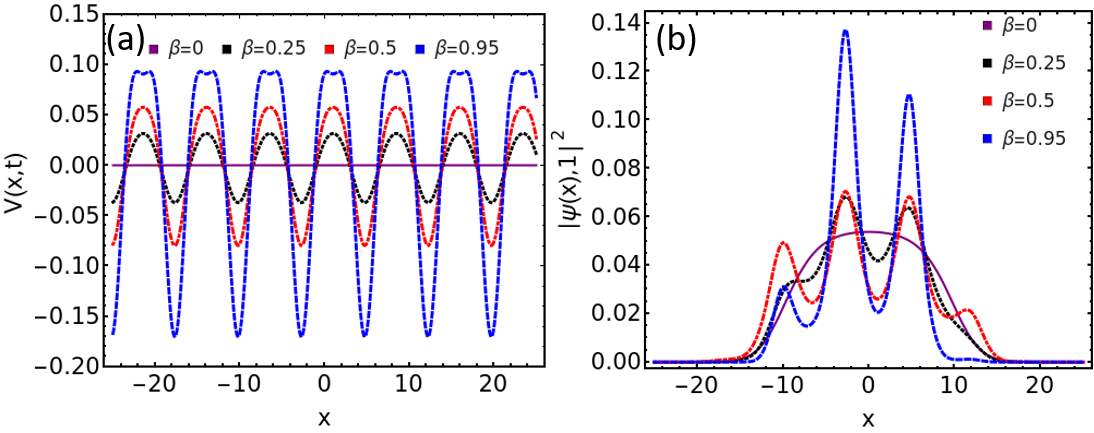}
\caption{\label{figI} External confinement and condensate density variation for a driven QOL trap with constant potential depth. As the magnitude of $\beta$ increases, the QOL potential depth deepens, leading to the formation of periodic structures in the QDs. The potential depth values are $\beta=0$ (purple), $\beta=0.25$ (black), $\beta=0.5$ (red), $\beta=0.95$ (blue). Other parameters are $k = 0.84$, $\mu=-2/9$, $G_{1}=-1$, $G_{2}=0.99999$, $v=1$, $t=1$. The spatial coordinate is scaled by the oscillator length.}
\end{figure*}
\begin{eqnarray}\label{eq:QD6}
[A^{2}(x,t)X_{x}(x,t)]_{x}=0, \\
X_{t}(x,t)+X_{x}(x,t)\phi_{x}(x,t)=0, \\ 
G_{1} X_{x}^{2}(x,t)-2 A(x,t) \gamma_{1}(x,t)=0, \\
G_{2} X_{x}^{2}(x,t)-2 A^{2}(x,t) \gamma_{2}(x,t)=0, \\
 \frac{A_{t}(x,t)}{A(x,t)}+\frac{1}{2 A^{2}(x,t}[A^{2}(x,t)\phi_{x}(x,t)]_{x}=0.
\end{eqnarray}
with
\begin{eqnarray}\label{eq:QD6a}
\hspace{-0.4 cm} \frac{A_{xx}(x,t)}{2A(x,t)}-\frac{\phi_{x}^{2}(x,t)}{2}-\phi_{t}(x,t)-\frac {1}{2} \mu X_{x}^{2}(x,t)-V(x,t)=0.\nonumber\\
\end{eqnarray}
In the equations above, a subscript denotes the partial derivative of the corresponding function with respect to the sub-scripted variable.  Here, $\mu$, represents the eigenvalue of Equation (\ref{eq:QD4}). The set of equations (\ref{eq:QD6})-(8) can be solved consistently to obtain:
\begin{eqnarray}\label{eq:QD7}
A(x,t)= \sqrt{\frac{c(t)}{X_{x}(x,t)}}, \;\;
\phi_{x,t}=-\frac {X_{t}(x,t)}{X_{x}(x,t)}, \\
\gamma_{1}(x,t) = G_{1} \frac{X_{x}^{5/2}(x,t)}{2 \sqrt{c(t)}}, \;\;
\gamma_{2}(x,t) = G_{2} \frac{X_{x}^{3}(x,t)}{2 c(t)},
\end{eqnarray}
where $c(t)$ is the constant of integration, and $G_1$, $G_2$ represents the the strength of BMF, and EMF interactions, respectively.  

From equation (\ref{eq:QD7}), it is evident that the amplitude, phase, and the EMF and BMF nonlinearities are directly governed by the function $X(x,t)$, which is determined by solving the consistency equation (\ref{eq:QD6a}). To achieve this, we substitute the trap expression from equation (\ref{eq:QD3}) into the consistency equation (\ref{eq:QD6a}) and define $X(x,t)= f[X(x,t)]=\int_{0}^{X} exp[\beta(t) \cos(k X)] \partial X$. This formulation allows us to derive the exact analytical expressions for the amplitude, phase, and nonlinearities:
\begin{eqnarray}\label{eq:QD7}
A(x,t)=\sqrt{\frac{1}{ exp[\beta(t) \cos(l X)]}}, \;\;
\phi(x,t)= \left[\frac{\beta^{2}(t)k^{2}}{16}-\frac{1}{2}v^{2}\right]t,\\
g_{1}(x,t)=\frac {G_{1} }{2} exp[\beta(t) \cos(l X)]^{\frac{5}{2}}, \;\;
g_{2}(x,t)=\frac {G_{2}}{2}exp[\beta(t) \cos(l X)]^{3},
\end{eqnarray}
such that $V_{1}(t)=-\frac{\beta^{2}(t) k^{2}}{16}, \;\;V_{2}(t)=\frac{\beta(t) k^{2}}{4},$  $X(x,t)=x-vt$,  and $c(t)=1$.

Finally, equation (\ref{eq:QD4}) represents the evolution of the droplets, for which an explicit solution can be formulated as: $U[X]=\frac{3 (\mu/G_{1}) }{1+\sqrt{1-\frac{\mu}{\mu_{0}} \frac{ G_{2}}{ G_{1}^{2}} } \cosh (\sqrt{-\mu}X)}$ with $\mu_{0}=-2/9$, $ E<0$, $G_1<0$, and $G_2>0$. \cite{Petrov,Astrakharchik}. Utilizing this and equation (\ref{eq:QD7}), we write the complete solution of the equation (\ref{eq:QD1}) as:
\begin{eqnarray}\label{eq:QD9}
\hspace{-2.8 cm} \psi(x,t)=\sqrt{\frac{1}{ exp[\beta(t) \cos\{k (x-v t)\}]}} 
\frac{\frac{3 \mu}{G_{1}} \times exp\left[i \{ \left[\frac{\beta^{2}(t)k^{2}}{16}-\frac{1}{2}v^{2}\right]t \} \right] }{1+\sqrt{1-\frac{\mu}{\mu_{0}} \frac{ G_{2}}{ G_{1}^{2}} } \cosh\left[\sqrt{-\mu} \int_{0}^{X(x,t)} exp[\beta(t) \cos \{k X(x,t)\}] \right]}, \nonumber\\
\end{eqnarray}
along with $\gamma_{1}(x,t)=\frac {G_{1} }{2} exp[\beta(t) \cos\{k (x- v t)\}]^{\frac{5}{2}}$,
$\gamma_{2}(x,t)=\frac {G_{2}}{2}exp[\beta(t) \cos\{k (x- v t)\}]^{3}$ . Here, $X(x,t)=x-vt$, $\mu_{0}=-2/9$, $ \mu<0$, $G_1<0$, and $G_2>0$. Thus, it is important to note that modifying the form of $\beta(t)$ enables the introduction of different temporal variations in the potential depth of the QOL. 

In the following section, we analyze the frequency spectrum of driven quantum droplets using FFT within the framework of the 1D eGPE, employing the exact analytical solution given by Eq. (\ref{eq:QD9}). The system is examined under the influence of temporally modulated repulsive cubic EMF and attractive quadratic BMF interactions. The droplet dynamics is studied by analyzing the frequency response of the system under varying driving velocities and potential depths of the QOL confinement. These investigations provide critical insights into the role of external modulation in the regulation of droplet stability and transport properties.

\begin{figure*}[t]
\centering
\includegraphics[width=\linewidth]{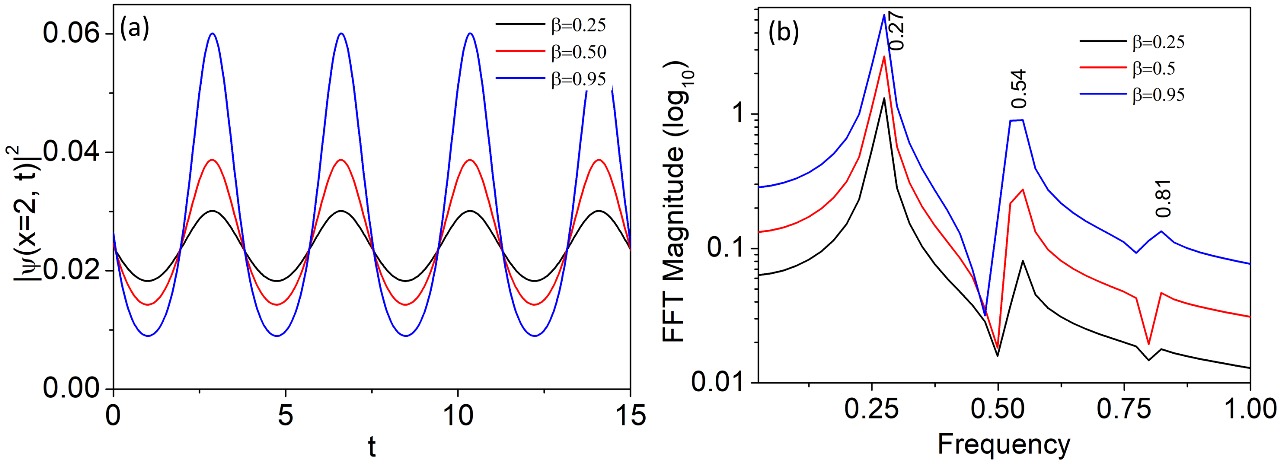}
\caption{\label{figII} (a) QDs density oscillations and (b) corresponding frequency spectrum for different values of $\beta$, with parameters $k = 0.84$, $v=2$, $\mu=-2/9$, $G_{1}=-1$, $G_{2}=0.99999$ and $x=2$. The frequency spikes for $\beta=0.25$, $\beta=0.5$ and $\beta=0.95$ are represented by black, red, and blue lines, respectively. The results indicate that while the overall frequency distribution remains similar, the spike heights increase with $\beta$. The spatial coordinate is scaled by the oscillator length.}
\end{figure*}

\section{The Condensate Dynamics and Frequency spectrum}
In this section, we investigate the dynamics of droplets under two experimentally relevant scenarios: (A) a time-independent QOL potential depth, $\beta(t)=\beta$, such that the resultant form of external trap becomes: $V(x,t) \simeq  -\frac{\beta^{2} k^{2}}{16} \cos[2 k (x-v t)] + \left[\frac{\beta k^{2}}{4} + \mu \beta \right] \cos[ k (x-v t)]$, and (B) a periodically modulated QOL potential depth, $\beta(t)=\beta_{0}(1+\alpha~cos(\omega t)$ undergoes sinusoidal oscillations at a constant frequency with $V(x,t) \simeq  -\frac{\beta^{2}(t) k^{2}}{16} \cos[2 k (x-v t)] + \left[\frac{\beta(t) k^{2}}{4} + \mu \beta(t) \right] \cos[ k (x-v t)]$. First, we examine the condensate density variation under the specified confinement conditions, followed by a FFT analysis to identify the resulting harmonic structures. The FFT, a numerical implementation of the discrete Fourier transform, decomposes a time-dependent signal into its frequency components, allowing us to extract key spectral features of the condensate dynamics. This approach enables a detailed characterization of oscillatory behaviors in the QDs system, providing insights into time crystal formation and the interplay between driving parameters, nonlinear interactions, and confinement effects.

\subsection{QOL confinement with time-independent potential depth} We consider the external confinement of the system in the form: 
\begin{equation}
V(x,t) \simeq  -\frac{\beta^{2} k^{2}}{16} \cos[2 k (x-v t)] + \left[\frac{\beta k^{2}}{4} + \mu \beta \right] \cos[ k (x-v t)],
\end{equation}
where the corresponding wavefunction follows from equation (\ref{eq:QD9}) with $\beta(t)=\beta$. The external confinement employed here is a bimodal drive characterized by two frequency components, \( f \) and \( 2f \), where \( f = kv/2\pi \). In figure (\ref{figI}) illustrates the profile of the chosen QOL confinement and the resulting condensate density at $t=1$ and $v=1$, as the potential depth is tuned via $\beta(t) = \beta$.  The considered values are $\beta=0$ (purple), $\beta=0.25$ (black), $\beta=0.5$ (red), $\beta=0.95$ (blue) with other parameters set as $k = 0.84$, $\mu=-2/9$, $G_{1}=-1$, $G_{2}=0.99999$, $v=1$, $t=1$. For $\beta=0$,  the system is effectively in free space ($V(x,t)=0$), resulting in the formation of a flat-top condensate density profile characteristic of a QD state \cite{Petrov}. As $\beta$ changes from $0 \rightarrow 0.95$  the potential depth of the QOL increases, leading to the fragmentation of the droplet profile into periodic droplet lattices, which eventually transition into bright solitons. Furthermore, increasing $\beta$ induces atomic density localization, a signature of disordered optical lattices such as QOL \cite{Maitri1,Maitri2}. The enhancement of $\beta$ introduces frustrated depths within the QOL trap, effectively lowering the potential barrier between adjacent lattice sites. This reduction in barrier height facilitates quantum tunneling of BEC atoms toward the trap center, reinforcing condensate density localization. The observed condensate density variation with increasing QOL potential depth closely resembles the fragmentation of QDs in optical lattices \cite{Ivan}. It is to be noted here that the lattice depth and frustration depth of the QOL trap are given by $(\beta +1)^{2}k^{2}/8$, and $(\beta -1)^{2} k^{2}/8$ respectively, with $\mu \rightarrow 0$. The corresponding location of maxima and minima are calculated as: $(2q +1) \cos^{-1}(\frac{1}{\beta})/k$ and $(q \pi)/k$, where $q$ is an integer \cite{Ajay1}. These expressions indicate that for $\beta \rightarrow 0$, the lower frequency component dominates, causing the system to effectively behave as a mono-frequency optical lattice, confirming the transition from a quasi-periodic to a regular lattice structure.

\begin{figure}[t]
\centering
\includegraphics[width=\linewidth]{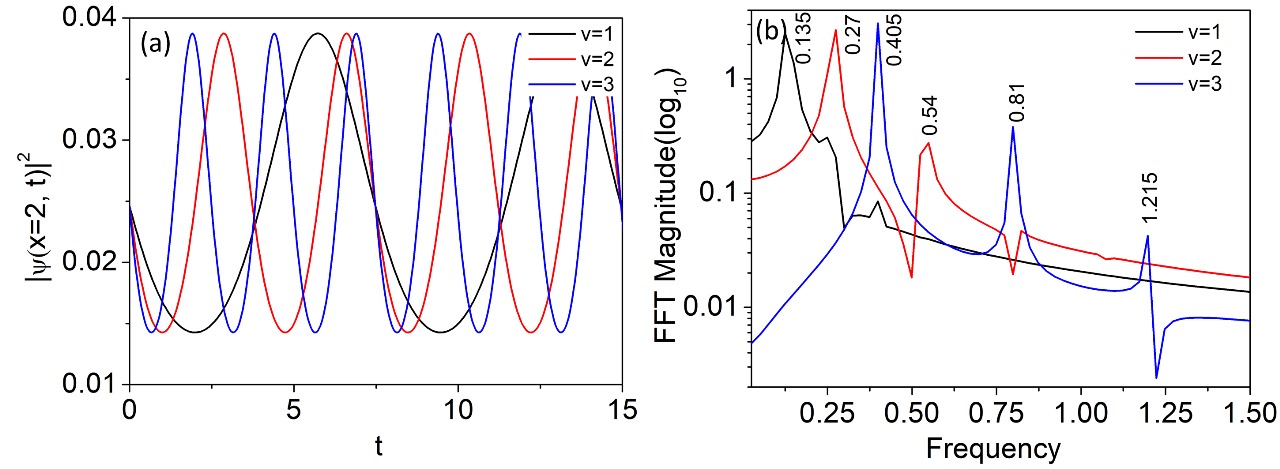}
\caption{\label{figIII} (a) QD density oscillations and (b) corresponding frequency spectrum for varying QOL driving velocity ($v$),  with parameters $k = 0.84, \beta=0.5, \mu=-2/9, G_{1}=-1, G_{2}=0.99999$ and $x=2$. The frequency spikes for $v= 1$ (black), $v=2$ (red) and $v=3$ (blue) show that, while the overall frequency distribution remains similar to Fig. \ref{figII}, the spikes shift to higher frequencies with increasing $v$. The spatial coordinate is scaled by the oscillator length.}
\end{figure}

We investigate two physical scenarios: (i) driving velocity ($v$) is constant along with potential depth ($\beta$) of QOL trap is increasing, and (ii)  potential depth ($\beta$) is constant and driving velocity ($v$) is increasing. First of all, in the figure \ref{figII} (a) and (b), we have illustrated the frequency spectrum corresponding to case (i). The density variation of the QD at $x=2$ is presented in figure \ref{figII} (a) and the corresponding frequency spectrum is presented in figure \ref{figII} (b). The variations of $\beta$ are $0.25$, $0.5$ and $0.95$ corresponding to the black, red, and blue line, respectively. Here, the parameters are $k = 0.84$, $v=2$, $\mu=-2/9$, $G_{1}=-1$, and $G_{2}=0.99999$, and the spatial coordinate is scaled by the oscillator length. As discussed earlier, for low values of $\beta$, the QOL trap transitions from a bichromatic to a monochromatic optical lattice. In the frequency spectrum, a secondary spike at $0.54$ appears alongside the dominant one at $0.27$. This secondary peak emerges as $\beta$ increases, indicating growing optical lattice frustration, which gradually transforms the monochromatic OL into a bichromatic QOL. The presence of this secondary spike serves as a clear signature of the emerging bichromaticity of the optical lattice.

\begin{figure}[t]
\centering
\includegraphics[width=\linewidth]{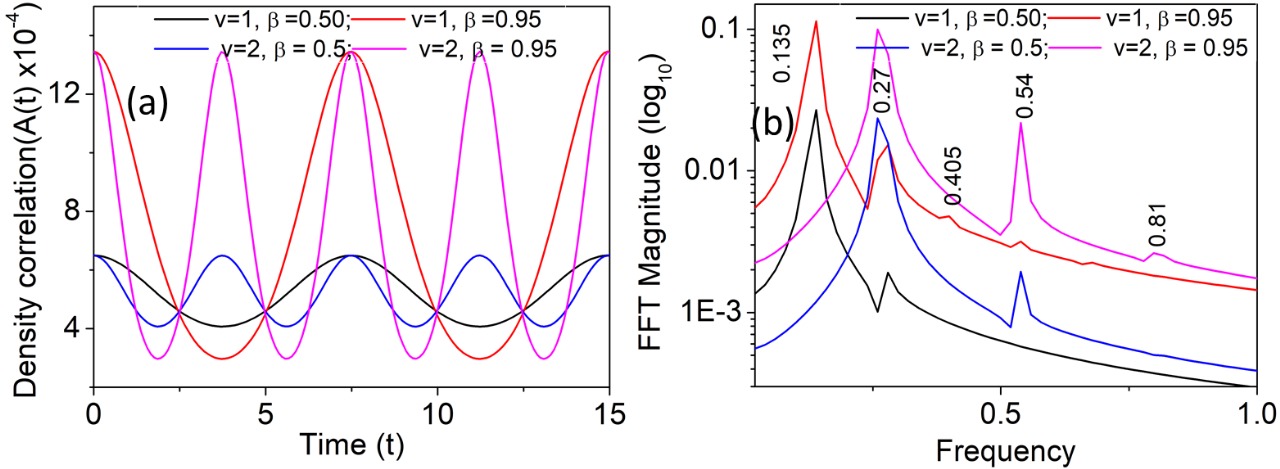}
\caption{\label{figIIIA} (a) Density-density auto-correlation function, and (b) corresponding frequency spectrum for QD density oscillations is plotted for four representative cases: (a) \( v = 1, \beta = 0.50 \) (black), (b) \( v = 2, \beta = 0.50 \) (blue), (c) \( v = 1, \beta = 0.95 \) (red), and (d) \( v = 2, \beta = 0.95 \) (pink). Here,  the parameters are $k = 0.84, \mu=-2/9, G_{1}=-1, G_{2}=0.99999$ and $x=2$. The frequency spikes for $v= 2$, and $\beta=0.95$ (pink) show that, along with $f$ and $2f$ driving frequency in FFT the $3f$ frequency observed to be present. The spatial coordinate is scaled by the oscillator length.}
\end{figure}

As shown in the FFT spectrum in Fig.~\ref{figII}(b), the QDs are driven by a fundamental frequency \( f = kv/2\pi \), along with its higher harmonics \( 2f \) and \( 3f \), where \( f \) depends on the driving velocity \( v \) and the QOL wavevector \( k \). For fixed \( v \) and \( k \), increasing the QOL depth parameter \( \beta \) enhances the oscillation amplitude of the droplet density, while the positions of the primary frequency peaks (at 0.27 and 0.54) remain unchanged. Notably, as \( \beta \) increases from 0.25 to 0.95, a new frequency component at \( 3f \) (i.e., 0.81) emerges, which is an integer multiple of the fundamental frequency and serves as a clear signature of discrete time-crystalline behavior. Thus, the system constitutes a periodically driven many-body quantum system that spontaneously breaks discrete time-translation symmetry, forming a discrete time crystal characterized by oscillations with a period that is an integer multiple of the driving period~\cite{Sacha1}. Moreover, the emergence of the discrete time crystal frequency \( 3f \) is closely tied to the depth of the QOL potential: for small values of \( \beta \rightarrow 0 \), corresponding to a nearly monochromatic periodic lattice, no such time-crystal frequencies are observed. These higher-order frequency components only appear beyond a threshold \( \beta \), highlighting the critical role of lattice-induced inhomogeneity in the formation of time-crystalline phases. 
 
\begin{figure}[t]
\centering
\includegraphics[width=\columnwidth]{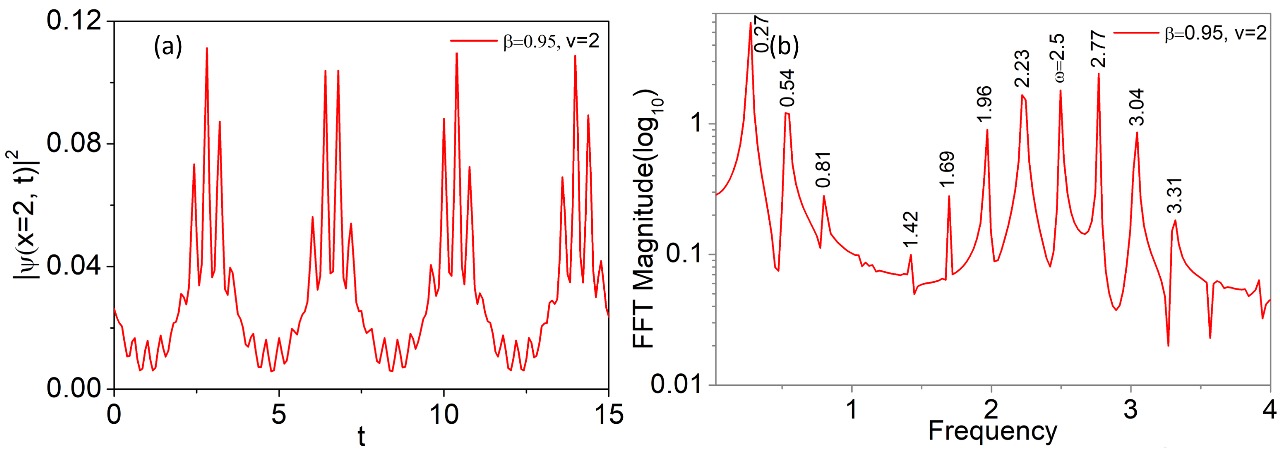}
\caption{\label{figIV} (a) QD density oscillations and (b) corresponding frequency spectrum when the QOL is moving with velocity $v=2$ and its depth oscillates sinusoidally with a constant frequency, given by $\beta(t)=\beta_{0}(1+\alpha~cos(\omega t)$. The parameters used are $k = 0.84$, $\beta_{0}=0.95$, $\mu=-2/9$, $G_{1}=-1$, $G_{2}=0.99999$, $\omega=2.5$ and $\alpha=0.6$ and $x=2$. A small perturbation in the QOL depth generates a combination of frequencies with the unperturbed frequency, labeled as: $f_{3}=3f_{1}=0.81$, $\omega - 3f_{1}=1.69$, $\omega - 2f_{1}=1.96$, $\omega - f_{1}=2.23$, $\omega =2.5$, $\omega + f_{1}=2.77$, $\omega + 2f_{1}= 3.04$, and $\omega + 3f_{1}=3.31$.}
\end{figure}
Next, we investigate the droplet dynamics for potential depth ($\beta$) is constant and driving velocity ($v$) is increasing for the same parameter values. In figure \ref{figIII} (a) and (b), we depict the density variation of the QD at $x=2$ and the corresponding frequency spectrum for increasing magnitude of QOL driving velocity from $1 \rightarrow 3$, respectively. The physical parameters are $k = 0.84$, $\beta=0.5$, $\mu=-2/9$, $G_{1}=-1$, and $G_{2}=0.99999$, and for the values of $v= 1$, $v=2$ and $v=3$ spikes presented with black, red and blue lines, respectively. For a moderate value of $\beta = 0.5$, where optical lattice OL frustration is present in a QOL, increasing the driving velocity of the chosen trap leads to droplet density oscillations similar to previous discussed case (i), except with a different periodicity. The frequency spike positions shift towards higher frequencies, specifically to $0.135$, $0.27$, and $0.39$ for $v=1$, $2$, and $3$, respectively. Here, the the quantum droplet oscillates with a constant frequency given by $3f$ in addition to the driving frequencies $f, 2f$ showing the signature presence of discrete time crystal in the chosen system. The presence of two spikes for a given $\beta$ and $v$ is due bichromaticity of QOL trap and the shift in spike positions occurs because increasing the OL velocity causes the QD to experience a rapid change in potential depth, resulting in faster oscillations.

Characterizing discrete time-crystalline order requires going beyond the analysis of local density modulations, as local observables alone may exhibit apparent frequency shifts arising from transient interference effects or externally driven mode mixing. A more definitive identification necessitates the examination of long-range temporal correlations to confirm the presence of robust, persistent subharmonic responses. To this end, we have performed a density–density auto-correlation function analysis, which serves as a sensitive diagnostic for identifying coherent time-crystalline behavior over extended evolution, defined as \cite{Pethick}: 
\begin{equation}
A(t) = \langle n(x,0) | n(x,t) \rangle = \int n(x,0)\, n(x,t)\, dx,
\end{equation}
where \( n(x,t) \) denotes the condensate density at time \( t \).

Figure~\ref{figIIIA}(a) shows \( A(t) \) for four representative cases: (a) \( v = 1, \beta = 0.50 \) (black), (b) \( v = 2, \beta = 0.50 \) (blue), (c) \( v = 1, \beta = 0.95 \) (red), and (d) \( v = 2, \beta = 0.95 \) (pink). For a fixed lattice velocity \( v \), increasing the QOL depth \( \beta \) leads to enhanced oscillation amplitudes in \( A(t) \) and a reduction in oscillation period, indicating stronger nonlinear coupling. The corresponding Fourier spectra, shown in Fig.~\ref{figIIIA}(b), exhibit not only the fundamental (\( f \)) and second harmonic (\( 2f \)) components, but also higher-order frequencies such as \( 3f \), which emerge and intensify with increasing \( \beta \). This growth of higher harmonics reflects the role of nonlinear interactions and lattice-induced inhomogeneity in facilitating subharmonic synchronization. In particular, the appearance and amplification of the \( 3f \) component with increasing QOL depth signify the formation and stabilization of a discrete time-crystalline phase. The use of the auto-correlation function \( A(t) \) provides a robust measure of long-range temporal coherence and has been widely employed as a diagnostic tool for identifying discrete time-translation symmetry breaking in driven many-body systems~\cite{Sacha2015, Else2016, Yao2017, Lledo2020}. 

\begin{figure}[t]
\centering
\includegraphics[width=\columnwidth]{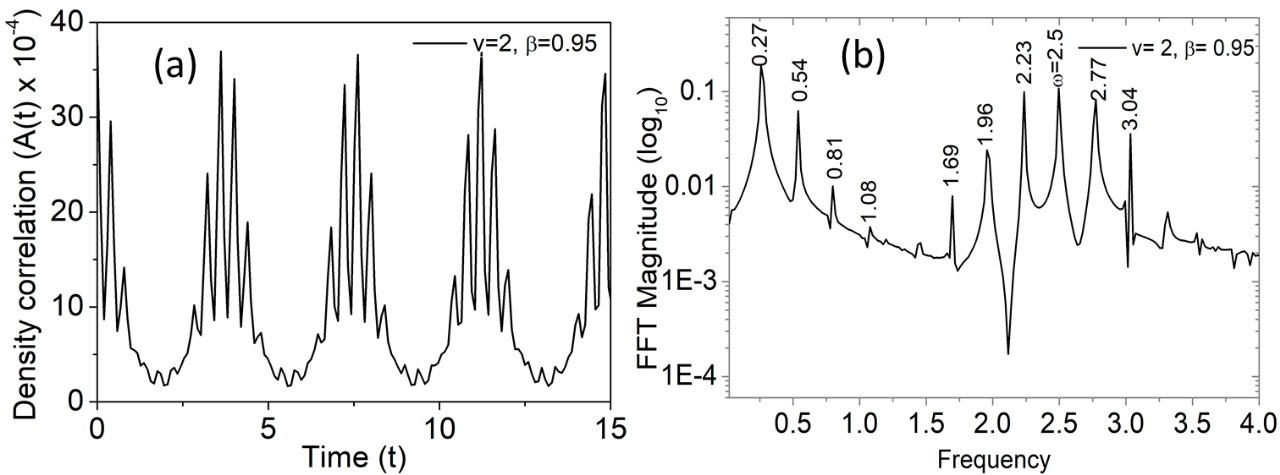}
\caption{\label{figIVa} (a), the temporal evolution of the density-density auto-correlation function \( A(t) \), and (b) corresponding FFT spectrum for \( \beta = 0.95 \) and \( v = 2 \), for $\beta(t)=\beta_{0}(1+\alpha~cos(\omega t)$. The parameters used are $k = 0.84$, $\beta_{0}=0.95$, $\mu=-2/9$, $G_{1}=-1$, $G_{2}=0.99999$, $\omega=2.5$ and $\alpha=0.6$ and $x=2$. A small perturbation in the QOL depth generates a combination of frequencies with the unperturbed frequency, labeled as: $f_{3}=3f_{1}=0.81$, $\omega - 3f_{1}=1.69$, $\omega - 2f_{1}=1.96$, $\omega - f_{1}=2.23$, $\omega =2.5$, $\omega + f_{1}=2.77$, $\omega + 2f_{1}= 3.04$, and $\omega + 3f_{1}=3.31$.}
\end{figure}

\subsection{QOL confinement with time-dependent potential depth} We now consider the scenario where the depth of the quasi-optical lattice (QOL) potential varies periodically in time and examine its impact on the emergence and characteristics of the time-crystalline frequency component $3f$. For that purpose, we choose$\beta(t)=\beta_{0}(1+\alpha~cos(\omega t)$, with $\omega$ is oscillating frequency and $\beta_{0}$, $\alpha$ are real constants. This results in the external confinement form: 

\begin{eqnarray}
\hspace{-2.4 cm} V(x,t) \simeq  -\frac{[\beta_{0}(1+\alpha~cos(\omega t)]]^{2} k^{2}}{16} \cos[2 k (x-v t)]  \nonumber\\
 + \left[\frac{[\beta_{0}(1+\alpha~cos(\omega t)]] k^{2}}{4} + \mu \beta \right] \cos[ k (x-v t)],   \nonumber\\  
\end{eqnarray}

for which the corresponding wavefunction is calculated from equation (\ref{eq:QD9}). It is important to note that the chosen external confinement incorporates three distinct frequencies: \( f \) (the frequency of the primary optical lattice), \( 2f \) (the frequency of the secondary lattice), and \( \omega \) (the temporal modulation frequency). As a result, the effective frequency spectrum of the potential \( V(x,t) \) includes components at \( f \), \( 2f \), \( \omega \), \( 2\omega \), as well as mixed terms such as \( \omega \pm f \) and \( 2\omega \pm 2f \).

\begin{figure*}[t]
\centering
\includegraphics[width=\linewidth]{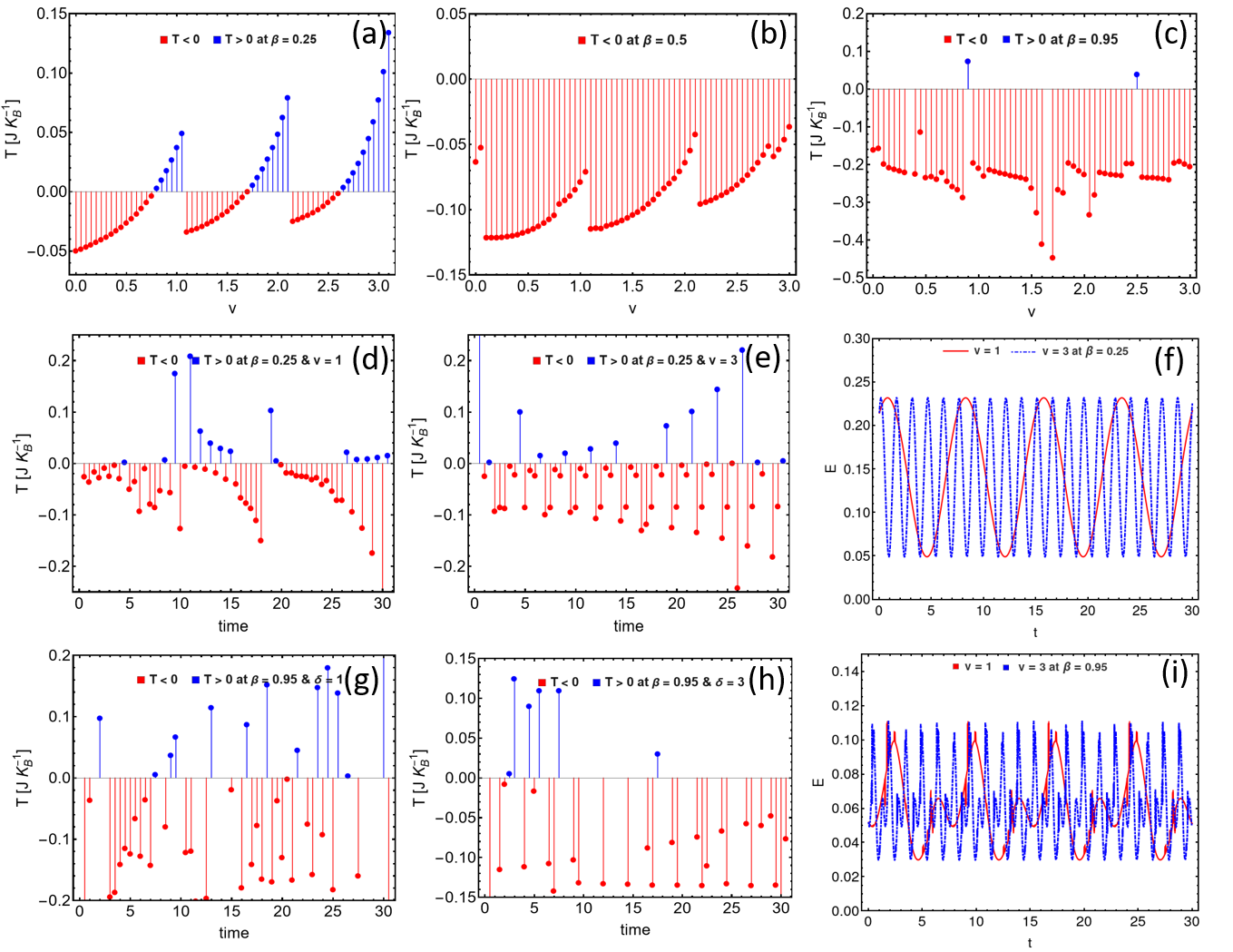}
\caption{\label{figV} This figure illustrates the variation of temperature ($T$) and total energy ($E$) with driving velocity ($v$) and QOL potential depth ($\beta(t) =\beta$). In panels (a)-(c), $\beta$ is varied from 0.25 to 0.95, while the driving velocity ranges from $v=0.01$ to $3$, depicting temperature evolution. Panels (d)-(f) and (g)-(i) show the time-dependent variation of temperature and total energy, respectively, for fixed $\beta = 0.25$, $0.95$. The parameters used are $k=0.84$, $\mu=-2/9$, $G_{1}=-1$, $G_{2}=0.99999$, with the spatial coordinate scaled by the oscillator length.}
\end{figure*}

In figure (\ref{figIV}), we present the condensate density evolution and its corresponding frequency spectrum for the representative case of \( v = 2 \). The system parameters are chosen as \( k = 0.84 \), \( \beta_{0} = 0.95 \), \( \mu = -2/9 \), \( G_{1} = -1 \), \( G_{2} = 0.99999 \), \( \omega = 2.5 \), and \( \alpha = 0.6 \). The presence of temporal modulation through \( \beta(t) = \beta_0(1 + \alpha \cos \omega t) \) enriches the frequency spectrum through both direct and mixed harmonic components. In addition to the fundamental (\( f_1 = 0.27 \)) and second harmonic (\( f_2 = 2f_1 = 0.54 \)) frequencies, the spectrum exhibits clear signatures of nonlinear mixing with the drive frequency \( \omega \), resulting in spectral components at:
\[
f_3 = 3f_1 = 0.81, \quad \omega - 3f_1 = 1.69, \quad \omega - 2f_1 = 1.96, \quad \omega - f_1 = 2.23,
\]
\[
\omega = 2.5, \quad \omega + f_1 = 2.77, \quad \omega + 2f_1 = 3.04, \quad \omega + 3f_1 = 3.31.
\]

These frequency components reflect the interplay of spatial translation and temporal modulation, leading to a rich spectral structure. Notably, the emergence of the \( 3f \) component is not present in the drive itself and thus serves as a hallmark of spontaneous subharmonic synchronization and discrete time-translation symmetry breaking. As illustrated in Fig.~\ref{figIVa}(a), the temporal evolution of the density-density auto-correlation function \( A(t) \) for \( \beta = 0.95 \) and \( v = 2 \) reveals coherent, long-lived oscillations. The corresponding frequency spectrum in Fig.~\ref{figIVa}(b) confirms the presence of fundamental components (\( f \) and \( 2f \)), drive-induced mixing terms (\( \omega \pm f \), \( 2\omega \pm 2f \)), and most importantly, a pronounced \( 3f \) peak. This frequency remains spectrally stable even in the presence of mixed harmonics, underscoring the \textit{robustness of the discrete time-crystalline phase} against temporal perturbations in the quasi-periodic optical lattice. The enhancement of the \( 3f \) peak with increasing lattice depth \( \beta \) further supports its \textit{nonlinear dynamical origin} and the formation of a stable time-crystalline order.

\section{Corelation between generated time crystal frequencies with negative temperature} In this section, we investigate the connection between the temperature and the time crystal frequencies generated in the driven QOL. According to the Kelvin definition of temperature, its inverse is given by the slope of entropy with respect to the system's energy \cite{Braun}. The system entropy is calculated as $S=-k_B\int_{-\infty}^{\infty}\rho(x,t)\ln\rho(x,t)dx dt$, where $k_B$ is the Boltzmann constant and $\rho$(x,t) is the condensate density \cite{Braun,Kundu}. The kinetic energy is determined using $E_k= \int_{-\infty}^{\infty}{|\frac{\partial\psi(x)}{\partial x}|}^{2}dx$. A negative temperature state arises when the entropy ($S$) is non-monotonic with kinetic energy ($E_k$) and reaches a maximum within the system's domain \cite{Braun}. Experimentally, Braun et al. demonstrated motional negative temperature states in weakly interacting $^{39}K$ BECs, and Kundu et al. illustrated that a bi-periodic optical lattice alone can induce negative temperature in a quasi-1D BEC, where its frustration depth serves as an upper energy limit \cite{Kundu}.

Based on these definitions, we estimate the system's temperature for two cases: (A) a time-independent QOL potential depth and (B) a periodically modulated QOL potential depth.

{\it (A). Time-independent QOL potential depth:} In figure (\ref{figV}), we represent the temperature variation by tuning the potential depth ($\beta$) and driving velocity ($v$) of the QOL confinement. In figure \ref{figV}(a)-(c), QOL potential depth parameter $\beta=0.25,0.5,0.95$, respectively and  driving velocity ($v$) is tuned from $0.01$ to $3$. It is apparent from the figure that with $\beta$ changing from $0.25 \rightarrow 0.95$ leads to increase in the magnitude negative temperature of the system with increasing $\beta$. It is attributed to the fact that increase in $\beta$ leads to increase in frustration depth of QOL trap and results in quantum tunneling of BEC atoms towards frustrated depth QOL. This ensures increased atomic occupation density at high energy frustrated depth resulting in decrease of negative temperature of the system \cite{Maitri2}. Further, in figures \ref{figV}(a)-(c), for a constant $\beta$, when $v$ is varied in between $0.01$ to $3$ leads to decrease in the magnitude of negative temperature of the system. 

\begin{figure*}[t]
\centering
\includegraphics[width=\linewidth]{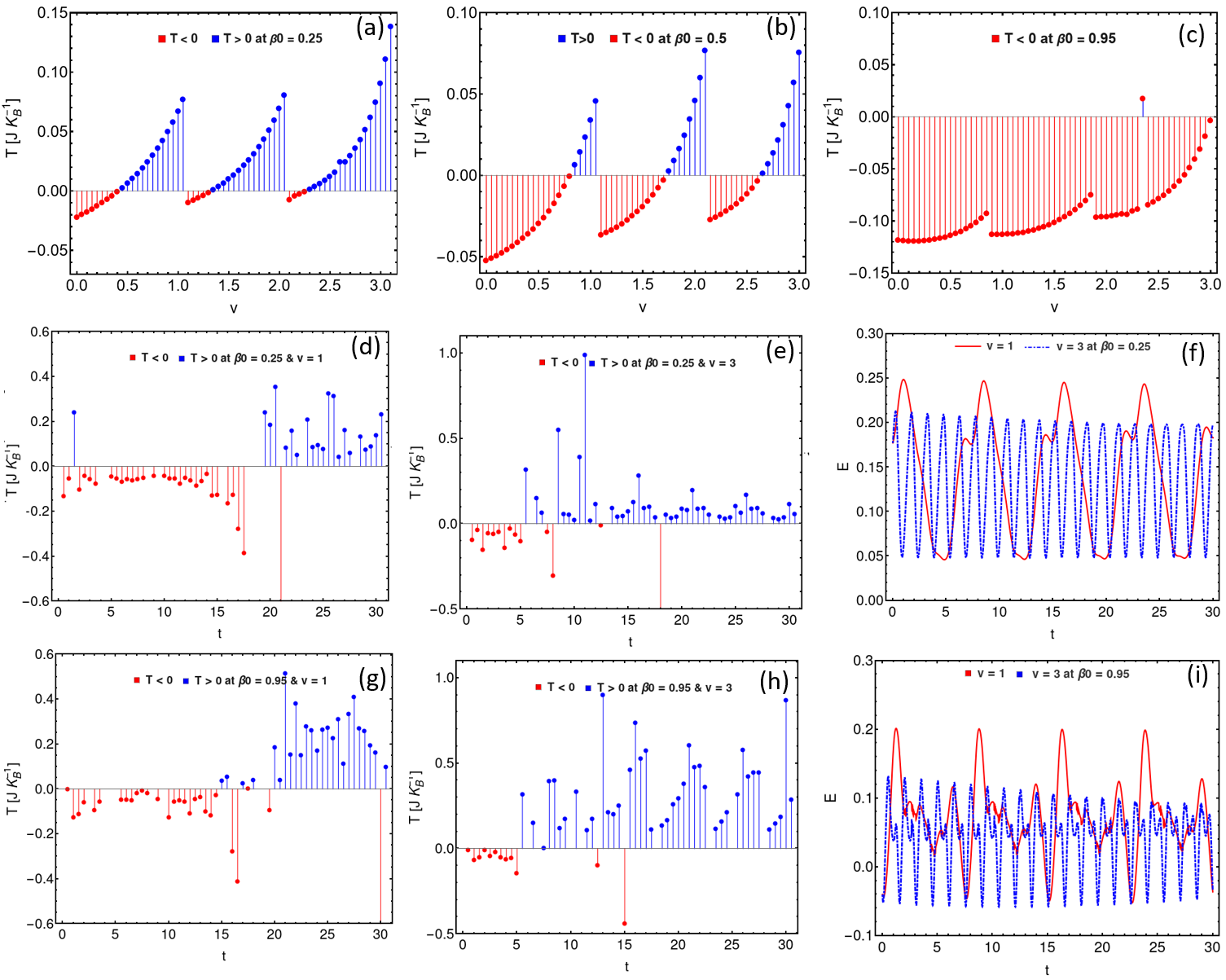}
\caption{\label{fig6} This figure illustrates the variation of temperature ($T$) and total energy ($E$) with driving velocity ($v$) and QOL potential depth [$\beta(t)=\beta_{0}(1+\alpha~cos(\omega t)$]. In panels (a)-(c), $\beta_{0}$ is varied from 0.25 to 0.95, while the driving velocity ranges from $v=0.01$ to $3$, depicting temperature evolution. Panels (d)-(f) and (g)-(i) show the time-dependent variation of temperature and total energy, respectively, for fixed $\beta_{0} = 0.25$, $0.95$. The parameters used are $\omega=2.5$, $\alpha=0.6$, $k=0.84$, $\mu=-2/9$, $G_{1}=-1$, $G_{2}=0.99999$, with the spatial coordinate scaled by the oscillator length.}
\end{figure*}

Additionally, in figures \ref{figV}[(d)-(f)] and \ref{figV}[(g)-(i)], we depict the variation of temperature ($T$) and total energy ($E$) of the system for $\beta=0.25$ with $v=1,3$ and $\beta=0.95$ with $v=1,3$, respectively. We calculate the total energy: $E = \int_{-\infty}^{+\infty} \left[ \frac{1}{2} \left(\frac{\partial \psi}{\partial x} \right)^{2} + \psi^{*} V(x,t) \psi -  g_1(x, t) |\psi| \psi + g_2(x, t)|\psi|^2 \psi \right]$. The increase in the magnitude of $v$ from $1 \rightarrow 3$ leads to increase in the periodicity of energy oscillation with time for $\beta =0.25$ in figure \ref{figV}(f). Further, the increase in $\beta$ from $0.25 \rightarrow$ i.e. increase in potential depth of QOL leads to quasi-periodicity in energy oscillation with time for both $v=1$ and $v=3$ velocities. In figure \ref{figV}[(d)-(e)] and figure \ref{figV}[(g)-(h)], the increase in $v$ from $1 \rightarrow 3$ leads to collapse of oscillation in the temperature variation with time for both $\beta =0.25$, and $0.95$. 

\begin{figure}[t]
\centering
\includegraphics[width=\linewidth]{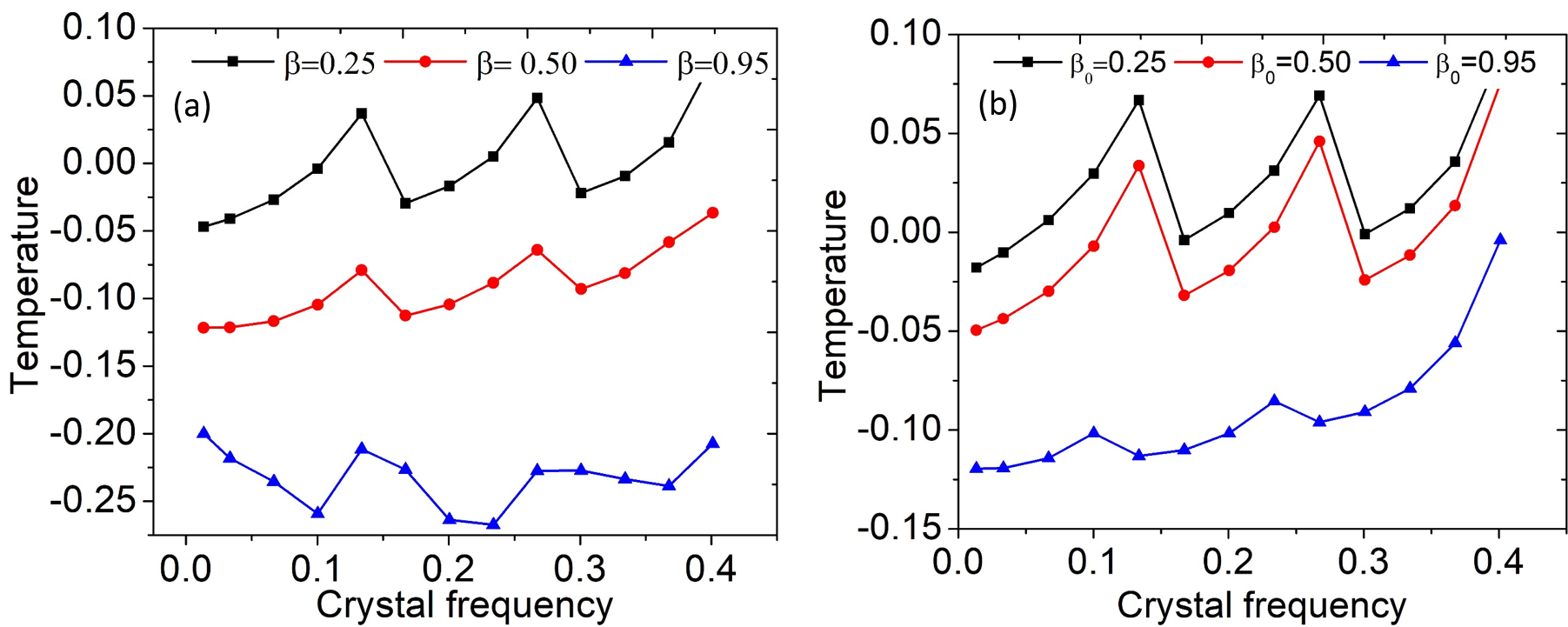}
\caption{\label{fig7} Temperature variation is illustrated with changing magnitude of time crystal frequency for (a) $\beta(t)=\beta$ for time-independent potential depth of QOL, and (b) time-dependent potential depth of QOL with $\beta(t)= =\beta_{0}(1+\alpha~cos(\omega t)$. The parameters used are $k=0.84$, $\mu=-0.22$, $G_{1}=-1$, $G_{2}=-0.9999$.}
\end{figure}

Further, in the figures (\ref{figII}) and ((\ref{figIII})), we illustrated that with $v$ changing from $1 \rightarrow 3$ leads to increase of discrete time crystal frequency by the relation $f=kv/2 \pi$ and at a $v$, the increase in $\beta$ leads to increase in size of time crystal frequency but it's position remains the same. Thus, the increase in the time crystal frequency (for $\beta$ at a constant magnitude and $v$ increasing) leads to decrease in the magnitude of negative temperature of the system whereas increase in the size of time crystal frequency (for increasing $\beta$ with constant $v$) leads to oscillatory nature in the magnitude of negative temperature of the system as depicted in figure \ref{fig7}(a). This analysis highlights that by tuning the parameters $\beta$ and $v$, one can effectively control the variations in both temperature (T) and total energy ($E$) providing insight into the interplay between the optical lattice depth, discrete time crystal formation, and thermodynamic properties of the system.

{\it (B). Time-dependent QOL potential depth:} Now, we investigate the variation of temperature and total energy variation for periodically oscillating potential depth of QOL with $\beta(t)=\beta_{0}(1+\alpha~cos(\omega t)$. Like in the previous case, in figure \ref{fig6}(a)-(c), QOL potential depth parameter is taken as $\beta_{0}=0.25,0.5,0.95$, respectively and  the driving velocity ($v$) is tuned from $0.01$ to $3$ for $\omega=2.5$, $\alpha=0.6$, $k=0.84$, $\mu=-2/9$, $G_{1}=-1$, $G_{2}=0.99999$. It is apparent from the figure that with $\beta_{0}$ changing from $0.25 \rightarrow 0.95$ leads to increase in the magnitude of negative temperature of the system with increasing $\beta_{0}$. As discussed above, it is attributed to the increased atomic occupation density at high energy frustrated depth resulting in increase of negative temperature of the system \cite{Maitri2}. Further, in figures \ref{fig6}(a)-(c), for a constant $\beta_{0}$, when $v$ is varied in between $0.01$ to $3$ leads to decrease in the magnitude of negative temperature of the system in all the three cases. In addition to that, increase in the magnitude of $v$ leads to decrease in magnitude of negative temperature and higher magnitude of $\beta_{0}$ suppresses the negative to positive temperature transition. 

In the figures \ref{fig6}[(d)-(f)] and \ref{fig6}[(g)-(i)] a transition in temperature from negative to positive is observed with increasing $v$ from $1 \rightarrow 3$ for both $\beta_{0}=0.25$ and $\beta_{0}=0.95$. This suggests that a higher driving velocity facilitates stronger energy exchange with the system, leading to an effective temperature shift. The variation of temperature ($T$) and total energy ($E$) of the system for $\beta_{0}=0.25$ with $v=1,3$ and $\beta_[0]=0.95$ with $v=1,3$ in the figures \ref{fig6}[(d)-(f)] and \ref{fig6}[(g)-(i)], respectively. With increase in magnitude of $v$ from $1 \rightarrow 3$ for $\beta_{0} = 0.25 \rightarrow 0.95$ leads to increase in the periodicity of oscillation of total energy in both cases. Additionally, for $v=3$ $\& \beta_{0}=0.95$ the emergence of beat formation is observed, indicating an interplay between multiple frequency components in the system. Furthermore, Fig. \ref{fig7}(b) illustrates the variation of temperature as a function of the time crystal frequency for different values of $\beta =0.25$, and $0.95$. It is evident that for a fixed $\beta_{0}$ an increase in time crystal frequency (achieved by increasing $v$) results in a oscillation in the magnitude of the negative temperature, indicating a oscillation of population inversion in the system. This highlights a nontrivial correlation between the optical lattice depth, the emergent time crystal frequency, and the thermodynamic properties of the system. This interplay suggests a deeper connection between the system’s nonequilibrium dynamics and its effective thermodynamic state, warranting further investigation.

\section{Numerical simulation and stability analysis:} In this section, we perform numerical simulations and stability analysis of the condensate density using equations (\ref{eq:QD1}) and (\ref{eq:QD9}) for the two previously mentioned cases, (A) and (B) of chosen QOL confinement. Specifically, we numerically solve equation (\ref{eq:QD1}) using the SSFT method in MATLAB. The results demonstrates strong consistency between numerical simulations and analytical solutions. Additionally, we analyze the stability of the QD solution under time evolution. 

\begin{figure}[t]
\centering
\includegraphics[width=\linewidth]{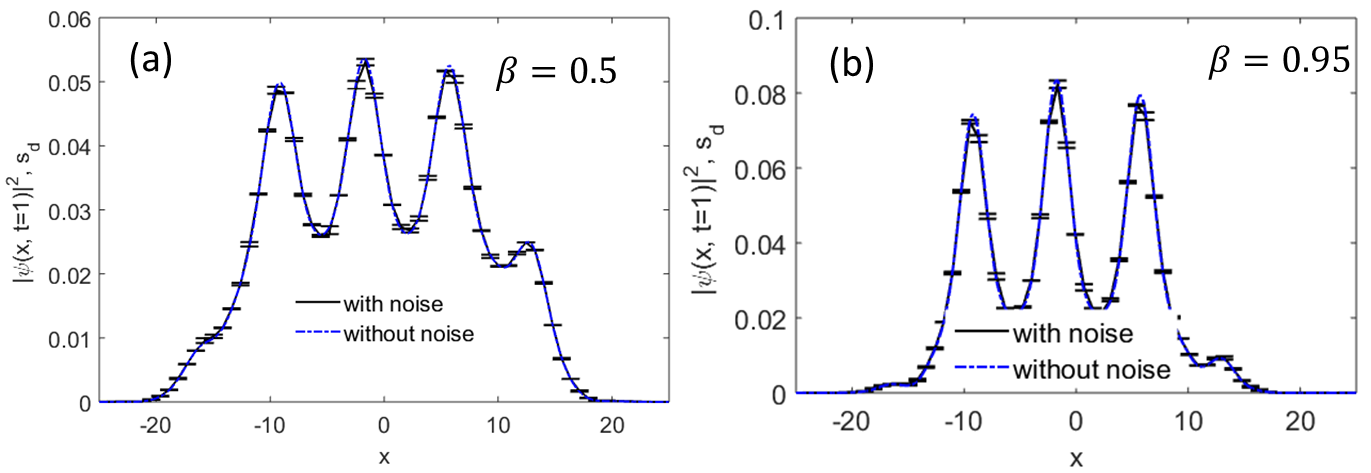}
\caption{\label{fig8} Structural stability and numerical simulation of quantum droplets (QDs). Panels (a) and (b) show the density standard deviation ($s_{d}$) for a noisy initial state (dashed blue line) and the numerically simulated density distribution after 10,000 iterations (solid black line) for $\beta=0.5$  and $\beta=0.95$, respectively. The parameters used are $k=0.84$, $\mu=-0.22$, $G_{1}=-1$, $G_{2}=-0.9999$ and $v=2$.}
\end{figure}

We consider two scenarios: (i) time evolution of the QD condensate under the given potential and (ii) stability against perturbations introduced by adding random white noise $R_{w}$ to the wavefunction. The noisy wavefunction is defined as: $\psi_{noisy}(x; t= 1) = \psi(x; t = 1) + R_{w}$. For the numerical simulations, we use equation (\ref{eq:QD4}) as the initial solution and evolve it under the potential given in Eq. (\ref{eq:QD3}) for $\beta=0.5$ and $0.95$. In the stability analysis, we take the noisy solution $\psi_{noisy}(x; t= 1)$   as the initial condition and evolve it in the same potential. The results indicate that, for the given noise $R_{w}$, the standard deviation ($s_{d}$) in the condensate density remains below $2\%$ of the density maximum, as shown in Fig. \ref{fig8}. This confirms that our analytical results are consistent with numerical simulations and that the obtained solutions are sufficiently stable for experimental applications. The stability analyses have been performed for the case when $\mu=-2/9$, $G_{1}=-1$, $G_{2}=-0.99999$, $k=0.84$ and $v=2$ for instance and is depicted in figure \ref{fig7}. The condensate density is observed for $10000$ time iterations with properly chosen space and time steps, $dx =0.078$ and $dt=0.0005$, respectively.

\section{Summary and Discussion} In this work, we have developed a comprehensive analytical framework to investigate the formation of time crystals and the dynamics of QDs in a driven QOL under negative temperature conditions. By deriving exact solutions to the 1D eGPE, we systematically analyzed the behavior of binary Bose-Einstein condensate systems under three distinct driving protocols. FFT analysis of condensate density profiles revealed the emergence of higher-order harmonics, thereby confirming the formation of time crystals—a hallmark of spontaneously broken discrete time-translation symmetry. Importantly, we have uncovered a non-trivial correlation between the frequency of the emergent time crystal modes and the effective thermodynamic temperature of the system. Specifically, we show that increasing the time crystal frequency leads to an oscillatory enhancement in the system’s negative temperature, offering new physical insights into the thermal behavior of out-of-equilibrium quantum matter. To substantiate our analytical predictions, we performed full numerical simulations and linear stability analysis using the split-step Fourier transform (SSFT) method. Our results demonstrate that the condensate density remains stable under temporal evolution and stochastic white noise perturbations, with density fluctuations confined within $2\%$ of the peak value. This agreement between analytical and numerical findings confirms the robustness of the proposed solutions and underscores their feasibility for experimental implementation. Overall, our study provides a foundational step toward understanding the interplay between time crystalline order and thermodynamic anomalies such as negative temperature in ultracold quantum systems. These results hold significant promise for advancing research in non-equilibrium many-body physics, quantum thermometry, and the design of quantum simulators and devices operating beyond conventional thermal regimes.

In this work, we report the emergence of discrete time-crystalline behavior in a 1D QDs governed by the eGPE under periodic driving via a QOL. Unlike quench-induced time crystals in undriven BECs, our system exhibits spontaneous subharmonic responses—including a robust \textit{3f} component—arising under continuous drive without reliance on special initial conditions or many-body localization mechanisms. Moreover, we distinguish our results from classical parametric resonance by demonstrating that the observed frequency mixing and period multiplication originate from nonlinear interaction effects intrinsic to the condensate dynamics. Using a density-density auto-correlation function analysis, we confirm the persistence of higher-order harmonic modes under both time-independent and time-dependent QOL depths, highlighting the role of lattice-induced inhomogeneity and disorder in symmetry breaking. While our approach is based on a coherent mean-field framework valid in the high-occupancy regime, we acknowledge that the eGPE does not capture thermalization or entropy growth. Our findings should thus be interpreted within the quasi-stationary coherent regime, consistent with prethermal time-crystalline behavior reported in related studies. Future investigations incorporating stochastic GPE or truncated Wigner methods will be essential to explore the impact of thermal fluctuations, decoherence, and long-time stability of the time-crystalline phase beyond the mean-field limit.

\section{Acknowledgment}

AN acknowledge insightful discussion with Prof. Prasanta K Panigrahi regarding the time crystal.

\section{Data availability}
All data generated or analysed during this study are included in this published article. It can be reproduced by utilizing the form of wavefunction and considered trap profiles.

\end{document}